\documentclass[10pt,conference]{IEEEtran} 

\AtBeginDocument{%
  \providecommand\BibTeX{{%
    \normalfont B\kern-0.5em{\scshape i\kern-0.25em b}\kern-0.8em\TeX}}}

\usepackage{multirow}
\usepackage{xcolor}
\usepackage{enumitem}
\usepackage{colortbl}
\usepackage{dblfloatfix}
\usepackage[width=0.9\textwidth,labelfont=bf]{caption}
\usepackage{pgfplots}
\usepackage{tikz}
\usepackage{tabu}
\usepackage{makecell}
\usepackage{xurl}
\usepackage{balance}
\usepackage{graphicx}
\pgfplotsset{compat=1.17}
\usepackage{array}
\newcolumntype{P}[1]{>{\centering\arraybackslash}p{#1}}

\definecolor{activeProject}{HTML}{C0504D}
\definecolor{baselineProject}{HTML}{9BBB59}

\begin{document}

%\title{Mining and Classifying GitHub Gender-Related Projects}
\title{Towards Understanding the Open Source Interest in Gender-Related GitHub Projects}

\author{
    \IEEEauthorblockN{Rita Garcia}
    \IEEEauthorblockA{Victoria University of Wellington\\
          Wellington, New Zealand\\
          rita.garcia@vuw.ac.nz}
    \and
    \IEEEauthorblockN{Christoph Treude}
    \IEEEauthorblockA{University of Melbourne\\
         Melbourne, Australia\\
         christoph.treude@unimelb.edu.au}
    \and 
       \IEEEauthorblockN{Wendy La}
    \IEEEauthorblockA{University of Adelaide\\
         Adelaide, Australia\\
         wendy.la@student.adelaide.edu.au}
 }

%\renewcommand{\shortauthors}{Anon, et al.}


\maketitle

\begin{abstract}

The open-source community uses the GitHub platform to exchange and share software applications and services of interest. This paper aims to identify the open-source community's interest in gender-related projects on GitHub. Our findings create research opportunities and identify resources by the open-source community that promote diversity, equity, and inclusion. We use data mining to identify GitHub projects that focus on gender-related topics. We apply quantitative and qualitative methodologies to examine the projects' attributes and to classify them within a gender social structure and a gender bias taxonomy. We aim to understand the open-source community's efforts and interests in gender topics through active projects. In this paper, we report on a preponderance of projects focusing on specific gender topics and identify those with a narrow focus. We examine projects focusing on gender bias and how they address this non-inclusive behaviour. Results show a propensity of GitHub projects focusing on recognising and detecting an individual's gender and a dearth of projects concentrating on the cultural expectations placed on women and men. In the gender bias domain, the projects mainly focus on occupational biases. These findings raise opportunities to address the limited focus of GitHub on gender-related topics through developing projects that mitigate exclusive behaviours.

\end{abstract}

\iffalse
\begin{CCSXML}
<ccs2012>
   <concept>
       <concept_id>10011007.10011006.10011072</concept_id>
       <concept_desc>Software and its engineering~Software libraries and repositories</concept_desc>
       <concept_significance>300</concept_significance>
       </concept>
   <concept>
       <concept_id>10003456.10010927.10003613</concept_id>
       <concept_desc>Social and professional topics~Gender</concept_desc>
       <concept_significance>500</concept_significance>
       </concept>
   <concept>
       <concept_id>10011007.10011006.10011071</concept_id>
       <concept_desc>Software and its engineering~Software configuration management and version control systems</concept_desc>
       <concept_significance>300</concept_significance>
       </concept>
 </ccs2012>
\end{CCSXML}

\ccsdesc[300]{Software and its engineering~Software libraries and repositories}
\ccsdesc[500]{Social and professional topics~Gender}
\ccsdesc[300]{Software and its engineering~Software configuration management and version control systems}
\fi

\begin{IEEEkeywords}
Gender; open-source community; GitHub projects; data mining
\end{IEEEkeywords}


\section{Introduction}

The open-source community uses GitHub to exchange and share software applications and services of interest to them. As of July 2022, GitHub had over 46 million public projects that can be used to report emerging technologies and trends across varying domains.\footnote{https://github.com/search?q=is:public} The paper's motivation is to identify the active gender-related projects by the open-source community on GitHub and report on a preponderance of projects focusing on specific gender topics and identify those topics with limited focus. We are interested in understanding the resources provided on GitHub by the open-source community that promote diversity, equity, and inclusion. We define \textit{gender-related projects} as projects that focus on sexual discrimination, gender bias, and stereotyping, along with feminism, health, family, and sexuality \cite{runyan:2014}. These topics appear within society's gender structure, influencing the roles and norms created by society for women and men \cite{risman:2013} at different dimensions, such as individual and institutional.

To find active gender-related projects, we use data mining, a technique previously used on GitHub to find information on the changing landscape for machine learning (ML) and artificial intelligence (AI) \cite{gonzalez:2020}. Data mining helped understand librarians' usage of the GitHub platform \cite{eaton:2021}. Analysing and reporting on GitHub projects can be done using data pipelines, tools that perform data processing to allow researchers and organisations to understand the collected data better \cite{munappy:2020}. Prior research has mined GitHub to understand the gender dynamics during software development, such as the gender-biased behaviours during Pull Requests (PRs), a submission process for software changes \cite{terrell:2017}. We use a pipeline to identify the open-source community's interests in gender-related topics through projects currently active on GitHub. We use the following research questions to guide our study:

\begin{itemize}
    \item \textbf{RQ1:} \textit{How do the attributes of the gender-related projects compare to other active GitHub projects?}
    \item \textbf{RQ2:} \textit{How do the gender-related projects align within the gender social structure?}
    \item \textbf{RQ3:} \textit{What types of gender bias are actively explored within GitHub projects?}
\end{itemize}

To answer the research questions, we apply a comparison study that uses quantitative and qualitative methods to analyse gender-related projects collected from a pipeline and compare them to a random selection of active GitHub projects to understand the open-source community's interests in this domain. The qualitative methods classify all gender-related projects within a gender social structure, helping to identify the concentration of GitHub projects within the gender social structure while opening future opportunities to address those areas with a narrow focus. We then align projects to a taxonomy focusing on gender bias, the conscious or unconscious awareness that influences a person's behaviour \cite{wang:2019} that can promote inequitable and non-inclusive environments. The open-source community might respond to gender bias through open-source services and applications, and we examine how they address this non-inclusive behaviour. To the best of our knowledge, this research is the first attempt to identify and report on the open-source community's interest in gender-related topics on the GitHub platform. The results show 235 active projects where most focus on recognising and detecting an individual's gender, with a slight interest in interactional behaviours, the cultural expectations placed on men and women \cite{risman:2013}. Within the gender bias domain, most of the projects address occupational bias, with discrimination having minimal representation.

%The show 10175 gender-related GitHub projects created over a ten-year (2011-2021) period, with , such as facial and gender recognition software applications. Our findings show missing metadata in GitHub projects, a situation also observed by \citet{kalliamvakou:2014}, making it difficult and unreliably to identify the purpose of problems. 

%Our findings echo prior research [] missing metadata, such as keywords, programming languages and labels, making it difficult for users to identify projects.

%In addition, our work contributed to modifying the systematic review pipeline to make it generic for others to apply it to other research topics. We presented the work performed on the data pipeline to apply to other research areas.

%\begin{itemize} %[leftmargin=*]
    %\item Describing the efforts and challenges to reusing an existing data pipeline when applied to a different domain.
%    \item Identifying active GitHub projects that focus on gender issues and topics. To the best of our knowledge, this is the first attempt to identify and report on gender-related projects.
%    \item Presenting the foci for the gender-related projects. We identified 235 active projects predominately focus on the detection and recognition of an individual's gender, with a small interest in interactional projects, projects that focus on the cultural expectations placed on men and women.
%\end{itemize}

The rest of the paper is structured as follows: Section \ref{section_related_work} presents the background. Section \ref{section_study_method} describes the research methodology. Section \ref{section_results} reports the results, while Section \ref{section_implications} presents the implications. Section \ref{section_threats_to_validity} describes the study's threats and concludes with Section~\ref{section_conclusion}, discussing future research opportunities from our work.

\section{Background} \label{section_related_work}

%This section presents the background in two sections: Section \ref{subsection_bkgd_mining} discusses mining projects on GitHub and Section \ref{subsection_bkgd_gender} presents gender-related research within the open-source community. 

\begin{figure*}[htp]
  \centering
  \includegraphics[width=.9\linewidth]{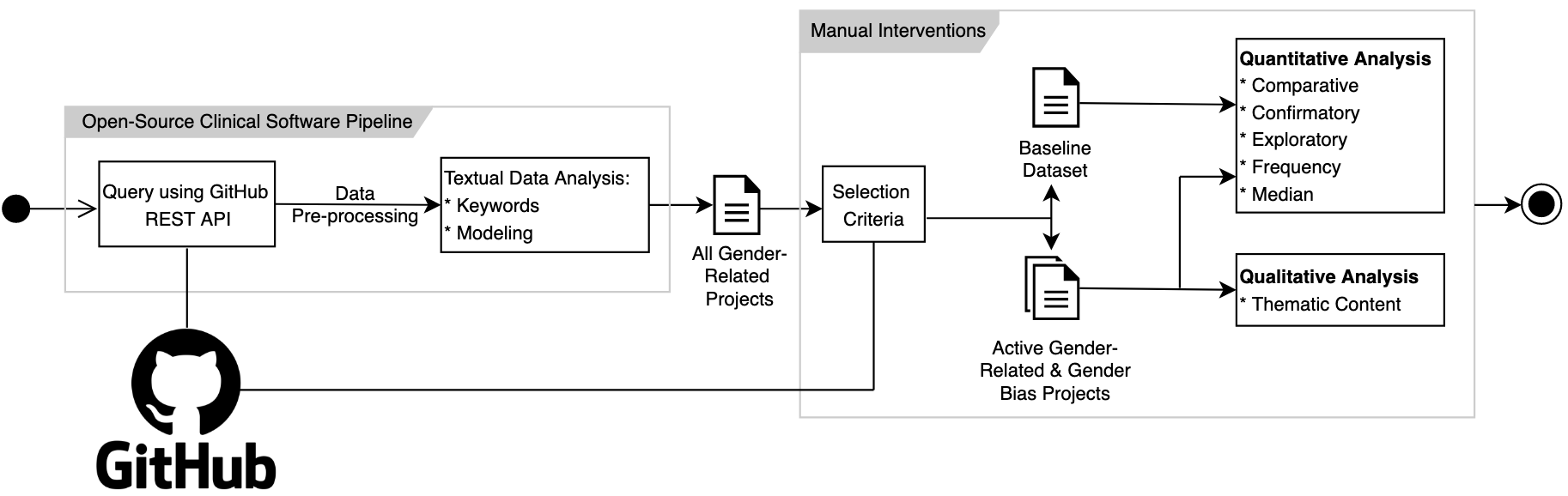}
  \vspace*{-6mm}\caption{Diagram of the Research Methodology}
  \label{fig:data_pipeline_figure}
\end{figure*}

\subsection{Mining GitHub Projects} \label{subsection_bkgd_mining}

%MSR enables researchers ``to support the maintenance of software systems, improve software design/reuse, and empirically validate novel ideas and techniques'' \cite[p.~89]{chaturvedi:2013}.

Mining Software Repositories (MSR) is a field of research that analyses data from software repositories and reports on the collected data \cite{herzig:2011}. Prior research has used MSR to examine gender diversity in software teams \cite{prana:2021}, identify the geolocations of GitHub contributors \cite{gousios:2014}, describe a Pull Development model \cite{gousios:2014}, and present the social features \cite{begel:2013} on GitHub. These exemplars demonstrate the diverse research opportunities from mining GitHub.

%Before Github, mining software was performed on SourceForge []. 
%Some of the previously mentioned research \cite{gousios:2014, gousios:2017} uses GHTorrent, using the collected timestamped data to mine software repositories. 

Technologies and tools are available to mine GitHub, such as the GitHub REST API used to create applications that extract data.\footnote{https://docs.github.com/en/rest} GHTorrent is a project that uses the GitHub REST API to collect, timestamp, and archive public projects.\footnote{https://ghtorrent.org} Like GHTorrent, GitHub Archive archives data hourly, making the data accessible via HTTP.\footnote{https://www.gharchive.org} Mining technologies can extract GitHub data, but additional work is required for interpretation since collected data can include biased results \cite{gousios:2017}, such as personal projects \cite{kalliamvakou:2014}. Avoidance strategies can identify active projects that contain a balance of commits, Pull Requests, and stars \cite{kalliamvakou:2014}. Stars represent interested users watching projects through update notifications \cite{zhang:2017}.

%Both GHTorrent and GH Archive provides historical data \cite{mombach:2018}. 
%since the findings ``can alter research conclusions if care is not taken to first establish that the data fits the research purpose'' \cite[p.~92]{kalliamvakou:2014}. 

 %A high number of project stars can signify a high-quality project []. 

%Query languages have also been developed to assist with mining GitHub projects. For example, Orion \cite{bissyande:2013} is a declarative query language that examines projects' metadata, their issue tracking systems, and source code to find relevant projects faster. Boa is another query language that provides users with a web interface to identify Java-based projects, reducing their programming efforts to find relevant projects \cite{dyer:2013}.

%MUDABlue also allowing the user to visualise the accurately categorised projects, along with providing the 
%Two techniques, CLAN (Closely reLated ApplicatioNs) and RepoPal, were evaluated by \cite{zhang:2017} to evaluate performance. 

%A comparison study \cite{zhang:2017} of these two techniques show RepoPal produced more accurate results than CLAN. 

%MUDABlue \cite{kawaguchi:2004} is a tool that identifies projects. Unlike RepoPal which uses metadata, MUDABlue examines the projects' source code, presenting the findings, such as their architectures and libraries, through a web interface.
%RepoPal is another technique that leverages projects' stars and \texttt{README} files to determine similarities. 

There are tools and techniques to mine projects. CLAN (Closely reLated ApplicatioNs) \cite{mcmillan:2012} is a technique that identifies similarities between Java projects through API method calls. A tool, \texttt{reaper} \cite{munaiah:2017}, identifies engineered software projects on GitHub. These projects leverage ``sound software engineering practices in one or more of its dimensions such as documentation, testing, and project management'' \cite[p.~3222]{munaiah:2017}. The \texttt{reaper} tool removes biased results, such as assignment projects, and identified 24.07\% of the 1,857,423 queried projects as engineered software projects. Frameworks support pipeline construction, like PyDriller \cite{spadini:2018}, a Python-based framework that reduces the complexity of mining projects but does not have features for presenting the findings. The tools, techniques, and mining approaches presented in this section provide researchers with various methods to construct pipelines for finding and reporting on extracted GitHub data. However, these approaches mainly focus on programming-related projects and our interests include software development projects and gender-related projects residing in other application domains, such as gender policy. 

%But through our research, we demonstrate how researchers can reuse existing pipelines for their data mining research. 

%Though these techniques find projects, additional work is required to analyse and report the findings.

%Seven dimensions were identified to establish engineered software projects, such as community, continuous integration, documentation, history, issues, license, and unit testing. Through the use of \texttt{reaper},

\vspace{-1mm}\subsection{Gender-Related Research in the Open-Source Community} \label{subsection_bkgd_gender}

%Because our research focus on gender issues and topics within GitHub, 
%We identify gender in OSS because it is well established that women are under-represented in the computing disciplines. In 2016 in the United States, 19\% of the CS Bachelor's degrees were awarded to women, a record decline from the 27\% awarded in 1997 \cite{nces:2016}.
%A report \cite{robles:2010} from 2016 shows that 2-5\% of the open-source participation comes from female contributors.

%The under-representation of women in Tech could be explained by the \textit{Social Construction Theory}. The theory argues socio-cultural factors influence the relationship people have with technologies \cite{marini:1990}, potentially contributing to the perception that Tech is a male domain.

We review the literature on gender-related topics within the open-source community to understand better how our work resides in this research area. One research area examines gender diversity in software teams, demonstrating women are under-represented as project contributors \cite{powell:2010}. A 2019 study \cite{bosu:2019} showed that 10\% of the GitHub project contributors were women, with 2.3\% comprising the core developers  \cite{canedo:2020}. This gender imbalance within teams can influence software products. For example, the under-representation of female developers within the FLOSS community has resulted in more unfriendly software towards women \cite{lin:2005}. Equal gender representation in software teams can bring together different perspectives and approaches to the development process; however, ``the software development profession does not reflect the people who use technology'' \cite[p.~20]{albusays:2021}. Gender diversity within teams can promote higher team performance \cite{rogelberg:1996}, helping to reduce \textit{community smells} \cite{catolino:2019} and the sub-optimal conditions within the software development communities \cite{tamburri:2015}. Diverse teams can handle issues well and resolve disagreements quickly \cite{earley:2000}, leading to higher team performance. A study \cite{vasilescu:2015} examining software commits showed diverse teams were more productive, concluding gender diversity can positively influence team productivity. To promote team diversity, Vasilescu et al. \cite{vasilescu:2015} suggest educational and professional incentives, such as training and outreach programs, to increase the presence of female contributors.

%using their software commits to measure productivity. The examination of 816 (24\%, 199 female) GitHub users' contributions on active projects showed the 

%Though there is support for having diversity in the software development process, there is gender differences in the open-source software development community, 

%A potential reason for the under-representation of women in the open-source community is the barriers they encounter within the community. 

Another research area examines causes for the under-representation of women in open source. Barriers, such as sexism and inequitable pay \cite{trinkenreich:2022}, challenge gender diversity and are a form of gender bias previously observed during the Pull Requests (PRs) process. PRs are source code changes that project maintainers review and comment on before acceptance \cite{kononenko:2018}. Terrell et al. \cite{terrell:2017} conducted a study that demonstrated when women's gender was identifiable during PRs, their requests were more likely to be rejected than male contributors. Another study by Imtiaz et al. \cite{imtiaz:2019} used a framework \cite{williams:2014} to examine contributors' GitHub PRs. The framework describes workplace patterns that influence women's career advancements. This study concluded that women concentrated their efforts on smaller sections of projects and a smaller group of organisations. Another study \cite{lee:2019} surveyed FLOSS contributors to collect their awareness of gender bias in the community, showing contributors' awareness of gender bias during software collaboration. Still, some contributors ``did not see the purpose of attempting to be inclusive, expressing that a discussion of gender has no place in FLOSS'' \cite[p.~677]{lee:2019}. Lee and Carver \cite{lee:2019} suggest a zero-tolerance policy for sexist behaviours to reduce challenges and make open source more inclusive. Women in the software industry recommend establishing role models and mentoring and creating inclusive events and groups to mitigate barriers \cite{trinkenreich:2022}.

The literature presented in this section examines gender disparity and bias women encounter in the open-source community. We reviewed literature that uses mining techniques on open-source projects to understand the diversity and dynamics in software teams. However, from reviewing the literature and to the best of our knowledge, our work is the first attempt to identify the open-source community’s interests in gender-related topics through open-source projects.

%s In a survey with Free-Libre/Open-Source Software (FLOSS) contributors across 15 projects, they were aware of gender bias in the community, such as sexism. 

%[ADD CONTENT] female contributors encounter with OSS culture \cite{holliger:2007}.

%Evaluating gender on GitHub has been performed in various areas on the platform, examining the diversity of contributors \cite{prana:2021} and the acceptance of Pull Requests by genders []. 

%Our literature review demonstrates prior work examining gender related topics, such as diversity and inclusion, in the OSS community, but prior research exploring gender on GitHub examined diversity with contributors. To the best of our knowledge, this is the first attempt examining GitHub projects that focus on gender issues and topics.

\section{Research Methodology} \label{section_study_method}

%This section presents the methods we used to report our findings. Section \ref{subsection_pipeline_design} describes the changes we made to the data pipeline, while Section \ref{subsection_data_collection} describes the how the information was extracted from the repositories. Section \ref{subsection_qual_analysis} describes the qualitative analysis performed on the projects that have high interest and software development.
 
We use a comparison study \cite{coccia:2018}, applying quantitative and qualitative methods to evaluate gender-related projects. Figure \ref{fig:data_pipeline_figure} presents an overview of how the data is collected (See Section \ref{section_pipeline_design}) and active projects selected (See Section \ref{subsection_selection_criteria}). Section \ref{subsection_quan_analysis} describes the quantitative analysis that compares attributes of the active gender-related projects with a random selection of active GitHub projects, indicating the level of interest in the gender-related projects. We refer to the random selection of the active GitHub projects as the \textit{baseline dataset} in Figure \ref{fig:data_pipeline_figure}. The qualitative analysis (See Section \ref{subsection_qual_analysis}) provides an in-depth view of the active GitHub projects. The comma-separated values (CSV) files discussed in this section are in our online appendix hosted on Figshare.\footnote{https://doi.org/10.6084/m9.figshare.19619034}

\begin{table}[h!]
\caption{Extracted Data Described by Shen and Spruit \cite{shen:2019} } \label{tbl:extracted_data}
\begin{tabular}{l|p{6cm}}
\textbf{Data Extracted} & \textbf{Description} \\ \hline

\multicolumn{2}{c}{\cellcolor[gray]{0.8}\textbf{Numerical Data Types}} \\ \hline
Contributors & The number of project contributors. \\ \hline
Creation Date & The project's creation date. \\ \hline
Forks & The number of project copies. Developers use forks for project contributions. \\ \hline
ID* & The project's unique key. \\ \hline
Issues & Number of open issues for the project. \\ \hline
\texttt{README} Size & The size of the project's \texttt{README} in KB. \\ \hline
Source Code Size & Total project size (KB), including its history. \\ \hline
Stars & The number of bookmarks for the project. \\ \hline
Update Date & The latest date the project was updated. \\ \hline

\multicolumn{2}{c}{\cellcolor[gray]{0.8}\textbf{Textual Data Types}} \\ \hline
Description & A brief explanation of the project.\\ 
\hline
Full Name* & Project's full name. \\ \hline
Languages &  Programming languages used in the project. \\ \hline
 Owner Location & Location of an owner. \\ \hline
 Owner Type & Type of owner: Individual or Organisation. \\ \hline
 \texttt{README} URL & URL to the project's \texttt{README} file.\\ \hline URL* & API URL providing project information. \\ 
 \multicolumn{2}{l}{* Data Not Described by Shen and Spruit \cite{shen:2019}}
\end{tabular}
\end{table}

%The pipeline supports our work in initially identifying gender-related GitHub projects. Because the pipeline helps in identifying all the gender-related projects, we apply the selection criteria, described in 

%to establish the active projects. This step is shown in Figure \ref{fig:data_pipeline_figure} and is the first manual intervention performed in the research methodology. The remaining manual interventions perform analysis of the collected data. 

\subsection{Data Pipeline} \label{section_pipeline_design}

We used a Python-based pipeline (See Figure \ref{fig:data_pipeline_figure}, ``Open-Source Clinical Software Project'') by Shen and Spruit \cite{shen:2019} that identifies open-source clinical software projects for reuse in the healthcare industry. This pipeline was selected for its extraction and analysis features, allowing us to mine GitHub projects within a different domain. Though other pipelines exist, such as a pipeline to identify Internet of Things (IoT) projects on the open web \cite{hwang:2016} and a Validation Pipeline \cite{hardle:2017} that evaluates YAML annotated software projects, these pipelines do not report the data necessary for this study's qualitative analysis (See Section \ref{subsection_qual_analysis}). 

%We used two scripts, where Figure \ref{fig:data_pipeline_figure} displays the scripts' features available in the scripts but displays the features and analysis processes we applied to our research. 

% and Plotly Chart Studio\footnote{https://plotly.com/python/chart-studio}

%and to save the generated files to a different directory on the local computer. 

%Reusing a data pipeline required setup, which included creating a token for the GitHub REST API and two accounts for the IBM-Watson and Plotly Chart Studio APIs. 

The pipeline uses the GitHub REST API to collect data, querying projects' ``name, description, \texttt{README} files, stars, forks, the number of contributors, and the number of commits'' \cite[p.~3]{shen:2019}. The pipeline provides the query as a string which we modified to \textit{``gender''}, identifying English-based projects created over ten years (2012-2021). The query results were saved as a CSV file, storing the projects' 17 attributes as separate columns. Table \ref{tbl:extracted_data} presents the attributes, organising them alphabetically within their data types: numerical and textual.

We examined the CSV file for data corruption that would impact our analysis and found two projects with corrupted data in the \textit{Description} field. Then, we used the pipeline for textual analysis on the \textit{Description} field to identify informative keywords. IBM-Watson was used for the textual analysis, saving the results to a CSV file for qualitative analysis (See Section \ref{subsection_qual_analysis}) to classify the projects.\footnote{https://pypi.org/project/ibm-watson}

\subsection{Selection Criteria for Active Projects} \label{subsection_selection_criteria}

%For our study, we seek to identify active gender-related projects, to better understand the open-source community's interest in developing products that focus on gender topics and issues. 

We developed the selection criteria to ensure we analysed valid projects and removed biased results, such as students' homework and dormant projects. The selection criteria select active projects with recent updates, commits, and Pull Requests. We applied recommendations by Kalliamvakou et al. \cite{kalliamvakou:2014} to construct the selection criteria considering nine perils to avoid biased results in findings. Table \ref{tbl:qual_selection_criteria} shows the four perils we used and how we applied them to identify active projects. We did not consider the five perils because they focus on software-based projects\textemdash~for example, \textit{Peril IV. A large portion of repositories is not for software development} \cite{kalliamvakou:2014}. These five perils may exclude projects from our research that do not require programming languages, such as those specialising in gender policy.

%, and software development processes, such as \textit{Perils VI}, \textit{VII}, and \textit{VIII} that focus on Pull Requests. We did not want to exclude projects that focus on gender policy or language. and we want to include these in our results.

\begin{table}[h!]
\caption{Selection Criteria for Active Projects} \label{tbl:qual_selection_criteria}
\begin{tabular}{p{3cm}|p{5cm}}
\hspace*{-1mm}\textbf{Peril} & \textbf{Selection Rule}\\
\hline
\hspace*{-1mm}A repository is not \newline \hfill \hspace*{-1mm}necessarily a project. & Select projects with three or more forks. \\ \hline 
\hspace*{-1mm}Most projects have \newline \hfill \hspace*{-1mm}very few commits. & \multirow{2}{*}{\parbox{4.6cm}{\vspace*{1mm}Select projects with updates, commits, or pull requests in the last six months (2021-08-01 to 2022-02-01).}} \\ \cline{1-1} % }} \\ \cline{1-3}
\hspace*{-1mm}Most projects are inactive.& \\ \hline 
\hspace*{-1mm}Two thirds of projects \newline \hfill \hspace*{-1mm}are personal. & Select projects with one or more stars. \\ \hline %because the data pipeline does not identify committers, we used stargazers to identify non-personal projects. 
\hspace*{-1mm}Most projects are inactive. & Select projects without \textit{``deprecated''}, \textit{``no longer maintained''}, \textit{``out-of-date''}, and \textit{``unsupported''} in the description field. %Any projects noted as deprecated are removed from the active projects.
\end{tabular}
\end{table}

We converted these perils into selection rules, shown in Table \ref{tbl:qual_selection_criteria}. For example, we addressed \textit{Peril III. Most projects are inactive} by selecting projects that demonstrate activity, such as commits, forks, and updates within six months (2021-08-01 to 2022-02-01). Six months was the duration because Kalliamvakou et al. \cite{kalliamvakou:2014} identified that 54\% of the GitHub projects were active during this period through commits and Pull Requests. We included project updates to the selection criteria because non-software projects might not use Pull Requests to make changes. The selection criteria also filtered dormant projects using a manual intervention similar to Tang et al. \cite{tang:2021} that used the keywords \textit{``deprecated''}, \textit{``no longer maintained''}, \textit{``out-of-date''}, and \textit{``unsupported''} to identify unsupported APIs in applications. In our study, we searched for the exact keywords in the projects' \textit{Description} field to exclude dormant projects. The selection process applied the selection rules in Table \ref{tbl:qual_selection_criteria} serially, which generated a CSV file with all active gender-related projects. We describe the analysis further in this section. 

%After identifying t￼ We used spreadsheets to perform median analysishe active projects, we analysed the projects using approaches described further in this section.

\subsection{Quantitative Analysis} \label{subsection_quan_analysis}

We used quantitative analysis to compare two datasets: the active gender-related projects and a baseline dataset containing a random selection of active GitHub projects. We construct the baseline dataset through random selection, a process previously used by Eaton \cite{eaton:2021} to create a baseline with GitHub users. But for our baseline, we randomly selected active projects using the selection criteria described in Section \ref{subsection_selection_criteria} and reviewed their descriptions to confirm they did not contain gender-related projects.

%\begin{enumerate}[leftmargin=*]
%    \item \textbf{All Gender-Related Projects on GitHub}: Data collected by the data pipeline.
%    \item \textbf{Active Gender-Related GitHub Projects}: See Section \ref{subsection_selection_criteria} for the creation of this dataset.
%    \item \textbf{Baseline Dataset}: A 
%\end{enumerate}

\begin{figure*}[t!]
  \centering
  \includegraphics[width=0.9\linewidth]{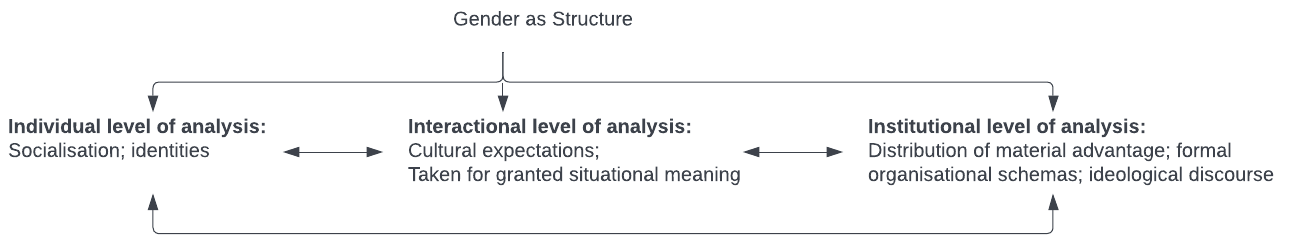}
  \caption{Gender Social Structure Framework \cite{risman:2013}} %to Classify Active Gender-Related Projects}
  \label{fig:risman_figure}
\end{figure*}

We used exploratory analysis to compare the projects' user types, forks, and stars. We selected these attributes because they provide a view into the open-source community's involvement (user type and forks) and interest (stars) in these projects. Exploratory analysis was previously applied to GitHub data, evaluating automotive projects \cite{kochanthara:2022} and examining the behaviours of librarians on GitHub \cite{eaton:2021}. The exploratory analysis was conducted on the two datasets in Excel, performing median analysis of the projects' user types, forks, and stars. 

Confirmatory analysis was used to determine the statistical significance of the gender-related projects. Pearson's chi-squared test (${\chi}^2$) was applied to the two datasets' user types. We also used two-sample t-tests to compare differences in the projects' forks and stars within the two datasets.

Comparative analysis was applied in this study to determine and compare the application domains used in the projects within the two datasets. This analysis provides a view into how the open-source community addresses gender-related topics using GitHub. Zanartu et al. \cite{zanartu:2022} designed a categorisation tool to identify open-source projects' application domains by processing their information using machine learning and natural language processing. The tool maps the projects across five application domains classified by Hudson et al. \cite{hudson:2016}. These application domains are \textit{Application \& System Software} that represents end-user systems, \textit{Documentation}, \textit{Non-Web Libraries \& Frameworks} that support the construction of non-web-based software applications and services, \textit{Software Tools} that support engineers in the development of software applications and services, and \textit{Web Libraries \& Frameworks} that support the construction of web-based software applications and services. The categorisation tool has a command-line interface that takes the projects' names as a CSV file and produces another CSV file containing the projects' application domains. After identifying the application domains, we performed frequency analysis in the Excel spreadsheets to create a distribution of application domains in the two datasets. 

%\begin{enumerate}[leftmargin=*]
%    \item \textbf{Application \& System Software:} Providing end-user systems, such as text editors.
%    \item \textbf{Documentation:} Focusing on documentation, such as coding exemplars and training materials.
%    \item \textbf{Non-Web Libraries \& Frameworks:} Providing frameworks and libraries to construct non-web-based software applications and services.
%    \item \textbf{Software Tools:} Providing tools that support engineers in the development of software applications and services, such as compilers and integrated development environments (IDEs).
%    \item \textbf{Web Libraries \& Frameworks:} Providing frameworks and libraries for web-based software applications and services.
%\end{enumerate}

%The Python-based categorisation tool processes the two datasets by importing the projects' names from their respective CSV files. 

We also used frequency analysis to evaluate the projects' programming languages. We identified the distribution of programming languages using a spreadsheet to compare the two datasets and to determine any differences in the languages used within these datasets. We performed the Mann-Whitney U Test \cite{greasley:2008} using IBM SPSS Statistics v25 to help identify any statistical differences between the projects' programming languages between the two datasets.

%which is provided by the project owners. Pre-processing the metadata was required to accurately quantify the data. For example, we adjusted the value \textit{``Python3''} to \textit{``Python''}. After pre-processing the programming languages, 

\subsection{Qualitative Analysis} \label{subsection_qual_analysis}

Qualitative analysis was used to classify the gender-related projects twofold. Section \ref{subsub_qual_all_projects} describes the first classification process, while Section \ref{subsub_qual_gb_projects} describes the second classification of projects focusing on gender bias.

\subsubsection{Classification of All Gender-Related Projects}\label{subsub_qual_all_projects}

We classified the active gender-related projects using thematic content analysis \cite{marshall:1999}. For the initial coding framework, we used a gender-social structure framework by Risman and Davis \cite{risman:2013} to help identify the concentration of projects within society's gender structure. The framework describes the roles and norms created by society for women and men within the social structure. Figure \ref{fig:risman_figure} shows the framework's three dimensions. The \textit{Individual} dimension focuses on the progress of gender identity and how the individual socialises. The \textit{Interactional} dimension focuses on the cultural expectations for women and men and the presumptions placed on them even when performing equivalent societal roles. The \textit{Institutional} dimension examines how culture influences actions through regulations and gender-specific organisational schemas. Risman and Davis posit the framework's dimensions have a cyclic relationship,  influencing the dynamics in the social structure that can contribute to society's gender inequalities.

%on two datasets constructed by the pipeline. The first dataset is the Google Scholar results. The data pipeline produced a spreadsheet of the results, including the keywords, publication title, and associated repository. For analysis, we identified the unique publications and identified publications that were associated with gender-related keywords. We analysed the keywords for these publications to identify common these in these papers. We strengthen the classification of these by also examining a subset of the collected repositories to identify the areas of interests the OSS community has on the gender-related topics. 

The initial coding framework contained four nodes: the three gender social structure dimensions and \textit{Unrelated} to represent projects unrelated to the gender topics. The pipeline extracted the projects' informative keywords (See Section \ref{section_pipeline_design}) to classify the projects into four nodes. The qualitative process marked any active project unrelated to gender into the \textit{Unrelated} node. The primary author performed the coding process with the projects' informative keywords. For coding reliability, the authors met to discuss the framework's dimensions to agree on the coding process. The meeting helped establish the accuracy of the coding process by confirming the authors' understanding of how the projects' keywords relate to the framework's dimensions. The authors decided to group \textit{``gender detection''} and \textit{``gender prediction''} keywords and place them within the \textit{Individual} dimension because the projects use the information to identify an individual's gender. The projects were re-coded after the discussion by one author to ensure the agreed-upon coding process was accurately applied to the projects. 

Because the projects contained a variety of informative keywords with similar meanings or goals, such as \textit{``Female Gender Classifier''} and \textit{``gender classifier''}, we used NVivo, a qualitative analysis tool, to consolidate similar keywords and phrases into groups. We used these groups to classify the projects into the framework's dimensions. If projects did not have informative keywords, we evaluated the \texttt{README}s to understand their purpose. Reviewing the \texttt{README} files was performed on 18.72\% of the projects exclusively using the keyword \textit{``gender''}, a broad term that lacks details on the project's purpose. During the classification process, we excluded project keywords (9.36\%) that did not relate to gender, such as \textit{``Javascript''} and \textit{``Must-read papers''}. If all the project's keywords were not gender-related, we placed the project into the \textit{Miscellaneous} group within the \textit{Unrelated} category. For projects (n=60, 20.34\%) containing multiple keywords related to gender, we coded them into more than one dimension. For example, the Inclusive Design Toolkit \cite{ontario:2022} is a project that promotes diversity and inclusion. This project was placed in the \textit{Inclusion} and \textit{Income} groups and classified within the \textit{Institutional} and \textit{Individual} dimensions. The coding process formed a matrix we extracted from NVivo to present the findings.

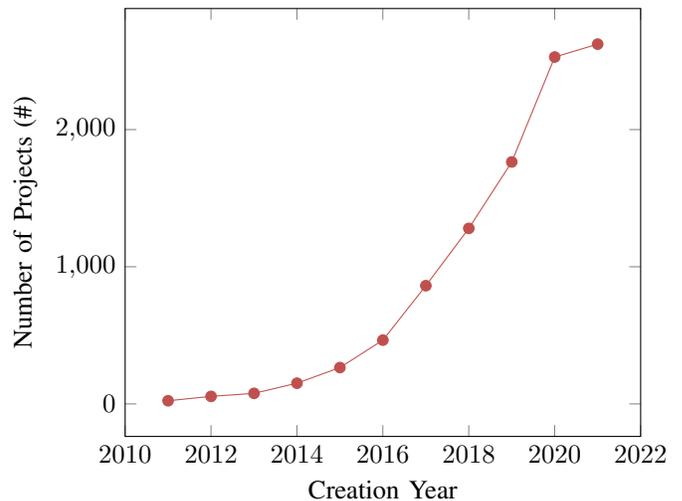
\begin{figure}[b!]
\centering
\begin{tikzpicture}
\begin{axis}[
xlabel=Creation Year,
ylabel=Number of Projects (\#),
x tick label style={/pgf/number format/.cd,%
          scaled x ticks = false,
          set thousands separator={},
          fixed},
legend pos=north west,
]
  \addplot[activeProject,mark=*] table [x=Year,y=Repos] {repoCreationData.txt};
\end{axis}
\end{tikzpicture}
\caption{Gender-Related Projects Cumulated by Creation Year} \label{results_new_repositories}
\end{figure}

\subsubsection{Classification of Gender Bias Projects}\label{subsub_qual_gb_projects}

%was used on the gender bias projects. \citet{doughman:2021} conducted their study on how Artificial Learning (AI), NLP, and ML have all reinforced the current gender biases and stereotypes. Their aim was to develop this taxonomy to aid in identifying these biases in language, specifically with English text. With several works around NLP, they found that although these works were able to minimise the bias, there was no connection between the detection and the societal significance and underlying application. 

We conducted the second classification after the gender-related projects were classified (See Section \ref{subsub_qual_all_projects}), involving the projects from the \textit{Gender Bias} group. We extracted these projects for thematic context analysis using a taxonomy by Doughman et al. \cite{doughman:2021} designed to help identify gender biases and stereotypes in the English language. This taxonomy was formed through observations on artificial intelligence, machine learning, and natural language processing, propagating and reinforcing prejudices when biased text is integrated into these systems. Table \ref{tbl:gender_bias_taxonomy_table} presents the five gender bias types in the taxonomy, along with the types' definitions. These five types are \textit{Generic Pronouns}, \textit{Semantics}, \textit{Sexism}, and \textit{Exclusionary} and \textit{Occupational Bias}. Each type contains subtypes. For example, the \textit{Sexism} type has \textit{Hostile} and \textit{Benevolent} sexism subtypes.

This classification process involved two authors to ensure coding reliability. The classification required inspecting multiple areas of the project to understand its purpose, such as source code. Using an Excel spreadsheet, the authors separately coded all the projects. After coding, the authors met to agree upon the coding criteria, which included a discussion of the discrepancies from the initial coding. The authors discussed their interpretations of the \textit{Exclusionary Bias} type and the subtypes within the \textit{Generic Pronouns} and \textit{Semantics}. The authors decided to apply \textit{Generic Pronouns} when masculine representation was assumed, while \textit{Semantics} was selected when addressing sexist stereotypes. The authors also negotiated emerging types, agreeing on the types and their labels. After the negotiation, one author re-coded all the gender bias projects using the agreed-upon coding criteria. Re-coding also included reviewing the projects' \texttt{README} files, artefacts, and any associated publications identified within the \texttt{README} files. Upon completion of the re-coding, the results were quantified in the spreadsheet.  

\begin{figure}[t!]
\begin{tikzpicture}
    \begin{axis}[
        width  = .4\textwidth,
        height = 4cm,
        x=.25\textwidth,
        major x tick style = transparent,
        enlarge x limits={abs=14mm},
        ybar=\pgflinewidth,
        bar width=4mm,
        ylabel = {Active Projects (\#)},
        symbolic x coords={Active Projects,Baseline Datasets},
        xtick = data,
        scaled y ticks = false,
        ymin=0,
        legend image code/.code={%
                    \draw[#1, draw=none] (0cm,-0.1cm) rectangle (0.6cm,0.1cm);
                },
        legend style={
         legend columns=1,
         legend cell align=left,
        at={(.60,.97)},
        draw=none,
        area legend,
    },
    ]
        \addplot[style={activeProject,fill=activeProject}]
            coordinates {(Active Projects, 195) (Baseline Datasets, 181)};

        \addplot[style={baselineProject,fill=baselineProject}]
             coordinates {(Active Projects,40) (Baseline Datasets, 54)};

        \legend{Individual, Organisation}
    \end{axis}
\end{tikzpicture}
\caption{Comparison of Projects' User Types} \label{fig:user_type_comparison}
\end{figure}
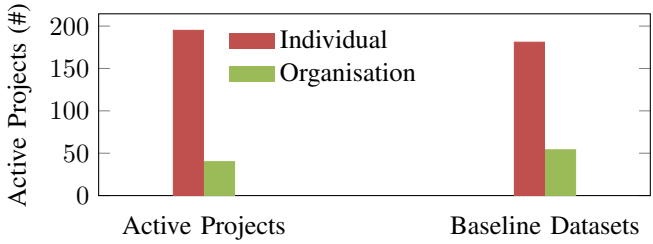

\section{Results and Discussion} \label{section_results}

%Before presenting the findings on the active gender-related projects, we provide an overview of all the gender-related projects in the past ten years (2011-2021). This information helps answer the research questions. All the gender-related projects can provide insight into users creating these projects and the open-source community's interest over the ten-year period. 

The pipeline's \textit{gender} query string identified 10,094 gender-related GitHub projects created over the ten years (2011-2021). Figure \ref{results_new_repositories} displays these projects by creation year, showing the number of projects increasing over the subsequent years. The query results show that 2017 to 2020 had over 400 newly created projects, with 2016 (n=397) approaching this figure. A potential reason for the increased number of projects using the term \textit{gender} might be the resurgence of the MeToo movement in 2017 \cite{mendes:2018}. This social movement raises awareness about sexual violence to let survivours know they are not alone. Though the movement was initiated in 2006 by Tarana Burke \cite{datla:2020}, the Twitter \#MeToo hashtag used by actress Alyssa Milano in October 2017 might have promoted interest within the digital space.

The selection criteria identified 235 (2.31\%) active gender-related projects from the 10,094 projects. The exclusion process removed 9661 projects with less than three forks, 160 projects without activity for six months, and 38 with zero stars. In this section, we further discuss the active gender-related projects by answering the paper’s three research questions. 

\subsection{RQ1: How do the attributes of the gender-related projects compare to other active GitHub projects?} \label{subsection_rq3}

%In this section, we compare these active projects to a baseline dataset representing a random selection of active projects on GitHub, evaluating the OSS community's interest (See Section \ref{subsub_user_type}) in the projects and the programming languages (See Section \ref{subsub_program_lang}) used in these projects.

%Five (2.12\%) of these active gender-related projects did not have keywords, which is closely aligned with keywords missing in the baseline dataset (X\%, n=X).

We compare the projects' attributes to understand how the 235 active gender-related projects align with a random selection of active GitHub projects. Figure \ref{fig:user_type_comparison} presents the comparison of the owner types. The table shows a similar composition of users creating gender-related projects and the random selection of active GitHub projects. The results show the majority (Active=82.98\%, Baseline=77.02\%) of the users are \textit{Individuals}, a term used by GitHub to differentiate projects owned by organisations and personal accounts. Our results are supported by Pearson's chi-squared test (${\chi}^2$ = 2.6064, \textit{p}-value = .106434) with a moderate effect size (\textit{p} $\leq$ 0.05) that showed no statistical differences between the datasets' user types.

%\subsubsection{Projects' Forks and Stars}

We also compared the projects' forks and stars attributes. Table \ref{tbl:p_test_comparison} shows a median value of 7 for the active gender-related projects' forks, while the median value for the baseline dataset projects is 44. Comparing the projects' stars also demonstrates similar results. There is a lower number (median=20) of stars for the active gender-related projects compared to the higher median value of 316 for the projects in the baseline dataset. The statistical significance supports these results in the forks ($t(235)_{Forks}$ = -4.05011, \textit{p} $<$ .00003) and stars ($t(235)_{Stars}$ = 9.07231, \textit{p} $<$ .00001) measured with moderate (.05) effect size. These results potentially demonstrate similar user types creating gender-related projects, but more work is required to raise awareness and interest within the open-source community about gender-related projects.

%The median for all gender-related projects' forks is 0 ($\overline{X}$=0.95), where 14.82\% (n=1508) contained more than one fork. The project with the most forks (1557 forks, 16.16\%) is a face classification project, a ``real time face detection and emotion/gender classification using fer 2013/imdb datasets with keras CNN model and openCV" \cite{arriaga:2022}. Reviewing the README for this project shows it has been depreciated and replaced with Perception for Autonomous Systems (PAZ) \cite{arriagapaz:2022}, a project that did not appear in our results because it does not directly reference gender. The median for the projects' stars is 0 ($\overline{X}$=2.85), where 23.66\% (n=2407) of the projects had one or more stars. The project with the most stars (17.84\%, n=5176) is also the face classification by Arriaga \cite{arriaga:2022}. 

\iffalse
\hspace*{-3mm}\fbox{
\begin{minipage}{.95\linewidth}

\textbf{Active Project User Types, Forks, and Stars Summary:}  The results show the projects are created in similar fashion as other active GitHub projects, similar interest by individuals and organisations. However, the open-source software community does not have similar interests to the gender-related projects as the average active projects on GitHub.

\end{minipage}}
\fi

%\subsubsection{Projects' Application Domains} \label{subsection_application_domain}

\begin{table}[t!]
%\vspace*{-8mm}
\caption{Comparison of Projects' Forks and Stars}\label{tbl:p_test_comparison}
\tabulinesep=0.5mm
\begin{tabular*}{\linewidth}{l|P{3cm}|P{2.8cm}}
%\begin{tabu}{l|c|c}
& \multicolumn{2}{c}{\textbf{Attributes ($\tilde{x}$)}} \\
\textbf{Datasets} & \textbf{Forks} & \textbf{Stars} \\ \hline
Gender-Related & 7 & 20 \\ \hline
Baseline & 44 & 316 \\ \hline
${\chi}^2$, \textit{p}-value &${\chi}^2$ = -4.050, \textit{p} $<$ .00003 &${\chi}^2$ = 9.072, \textit{p} $<$ .00001 \\
\multicolumn{3}{l}{The result is significant at p $<$ .05.}
\end{tabular*}
\end{table}

%USE MAKECELL \makecell{${\chi}^2$ = 9.072, \\  \textit{p} $<$ .00001} \\

\begin{table}[b!]
\caption{Comparison of Projects' Application Domains} \label{tbl:application_domain_comparison}
\begin{tabular*}{\linewidth}{@{\extracolsep{\fill}}p{0.5\linewidth}|p{0.2\linewidth}|p{0.2\linewidth}@{}}
\textbf{Domain} & \textbf{All Gender}&\textbf{Baseline}\\
\hline
Application \& System Software & 7 (2.98\%) & 9 (3.83\%) \\
Documentation&	34 (14.47\%) & 16 (6.81\%) \\
Non-Web Libraries \& Frameworks & 166 (70.64\%) & 72 (30.64\%) \\
Software Tools & 8 (3.40\%) & 44 (18.72\%) \\
Web Libraries \& Frameworks& 20 (8.51\%) & 94 (40.00\%) \\
\end{tabular*}
\end{table}

\begin{figure*}[b!]
\begin{tikzpicture}
    \begin{axis}[
        width  = \textwidth,
        height = 5cm,
        major x tick style = transparent,
        ybar=3*\pgflinewidth,
        bar width=14pt,
        ytick={0,20,40,60,80,100},
        ylabel = {Active Projects (\#)},
        symbolic x coords={C++,HTML,Java,JavaScript,Jupyter NB,None,PHP,Python,R,Ruby},
        xtick = data,
        scaled y ticks = false,
        ymin=0,
        legend image code/.code={%
                    \draw[#1, draw=none] (0cm,-0.1cm) rectangle (0.6cm,0.1cm);
                },
        legend style={
         legend columns=1,
        at={(.35,.97)},
        draw=none,
        legend cell align=left,
        area legend,
    },
    ]
        \addplot[style={activeProject,fill=activeProject,mark=none}]
            coordinates {(C++, 8) (HTML, 9) (Java,7) (JavaScript,11) (Jupyter NB, 42) (None, 12) (PHP, 9) (Python, 101) (R, 12) (Ruby, 4)};

        \addplot[style={baselineProject,fill=baselineProject,mark=none}]
             coordinates {(C++,0) (HTML, 6) (Java,21) (JavaScript, 92) (Jupyter NB, 0) (None,6) (PHP, 12) (Python, 21) (R, 0) (Ruby, 17)};

        \legend{Active Gender-Related Projects, Baseline Dataset}
    \end{axis}
\end{tikzpicture}
\vspace{-3mm}\caption{Comparison of Projects' Programming Languages} \label{fig:comparison_programming_languages}
\end{figure*}
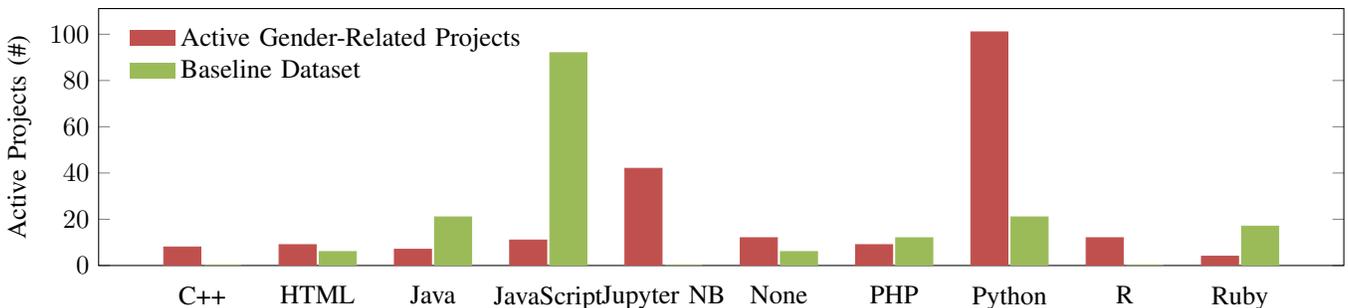

Results from the application domains provide insight into the types of projects created. Table \ref{tbl:application_domain_comparison} shows the \textit{Non-Web Libraries \& Frameworks} application domain is the most prominent (n=166, 70.64\%) within the active gender-related projects. In contrast, the random selection of GitHub projects in the baseline dataset focused on the \textit{Web Libraries \& Framework} application domain (n=94, 40.00\%). Both datasets show limited focus on the \textit{Application \& System Software} application domain, where the active gender-related projects had seven (2.98\%) and nine (3.83\%) projects in the baseline dataset. A potential reason for these findings might be that gender-related projects contain task-based scripts, such as automated gender detection. The gender problem domain might not suit the solutions presented as stand-alone applications.

Figure \ref{fig:comparison_programming_languages} compares the top-10 programming languages used by the gender-related projects and the baseline dataset. The results show Python (n=101, 45.29\%) was predominately used by the active gender-related projects, while the random selection of active GitHub projects predominately used JavaScript (n=92, 39.15\%). The Mann-Whitney U Tests (\textit{p}$<$0.05 two-tailed) show significant differences for the \textit{JavaScript}, \textit{Jupyter Notebook}, and \textit{Python} languages when comparing the two datasets. More research is required to understand the different applications of languages in gender-related projects.

%\subsubsection{Projects' Programming Languages} \label{subsub_program_lang}

%With the majority (94.89\%) of the active projects providing programming languages, 

%Of the 235 active projects, 12 (5.11\%) did not declare a programming language. 

%into two different classifications: \textit{``gender detection''} and \textit{``gender prediction''}. 

%\vspace*{-3mm}
\hspace*{-4mm}\fbox{
\begin{minipage}{.97\linewidth}

\textbf{RQ1 Summary:}  The active gender-related projects have a similar composition of project owners as a random selection of active GitHub projects; however, the forks and stars results show lower interest by the open-source community in these gender-relate projects. When comparing the two datasets, we found differences in the programming languages used by the projects, but both have limited representation within \textit{Application \& System Software} application domain. Work is required to understand the application scenarios for gender-related projects and how to promote interest in these projects within the open-source community.

%When reporting on the data pipeline output, we report on how the information provided by the project maintainers influenced the results. Areas with limited provided data include the repositories \textit{keywords}, and \textit{programming languages}, potentially making it difficult for users to find projects appropriate for their work. Our findings echo the prior research by \cite{shen:2019} that showed lack of data provided by project maintainers in the metadata. Therefore, we encourage project maintainers to provide information metadata. Presenting the results from the data pipeline shows more accurate information can be provided with additional information manually added by project maintainers. We recommend, like the clinical software repositories, provide more descriptive metadata. With additional information, users might be able to find repositories that are appropriate for their work.

\end{minipage}}

\subsection{RQ2: How do the active gender-related GitHub projects align within the gender social structure?} \label{subsection_rq4}

We classified 235 active projects within the gender social structure framework \cite{risman:2013}, shown in Table \ref{tbl:risman_gender_results}. The table organises the findings within the three gender social structure dimensions and an \textit{Unrelated} category in descending order. We used the projects' informative keywords for the coding process, but we also evaluated 71 (30.71\%) projects' \texttt{README} files to understand their purpose better. Within these 71 projects, 44 (18.72\%) used the keyword \textit{``gender''}, 22 projects (9.36\%) contained only keywords unrelated to gender, such as \textit{``R package''} and \textit{``Chrome extension''}, and five (2.13\%) projects required further analysis because they did not have keywords. As a result, 25 (8.47\%) projects were placed in the \textit{Unrelated} category to represent projects that could not be classified within the gender social structure. 

%The coding results produced 16 project groups within these four categories, where the projects were represented as 295 nodes. 

Table \ref{tbl:risman_gender_results} shows the coding results with 16 project groups across four categories. The table organises the groups in descending order within the categories, providing group descriptions and an exemplar project as an illustration. For example, the \textit{Gender Diversity} group resides in the \textit{Institutional} (n=32, 10.85\%) dimension and contains seven (2.37\%) projects.

%Further inspection of these projects included examining their documentation, such as \texttt{README} files. 
%295 total classification
\begin{table*}[thp!]
\caption{Classification of Active GitHub Projects within the Gender Social Structure Framework} \label{tbl:risman_gender_results}
\begin{tabular}{p{1.6cm}|c|p{5.4cm}|p{7.9cm}}
\textbf{Group} & \textbf{Total (\%)} & \textbf{Project Descriptions}& \textbf{Example Project}\\ \hline
\multicolumn{4}{c}{\cellcolor[gray]{0.8}\textbf{Individual (n=219, 74.24\%)}} \\ \hline
Detectors and \newline \hfill Predictors & 164 (55.59\%)& Detects faces and emotions using age and gender; Uses facial attributes to estimate age and gender; Trains ML models to detect gender; Detects gender, race, and emotion in real-time. & FaceLib \cite{ayoubi:2022} is a Python-based library that recognises facial expressions and gender using an image file of an individual. \\ \hline
%Gender Recognition & 36 (12.20\%) & Recognises gender, race, and emotion, some in real-time. \\ \hline
Gender of a \newline \hfill Name & 25 (8.47\%) & Assists in legal name changes; Guesses and trains models to detect gender from names. & \texttt{randomNames} \cite{betebenner:2022} contains a function that randomly selects names based on gender as the input. \\ \hline
Gender \newline \hfill Classification & 20 (6.78\%) & Classifies speakers gender using names, voices, and images. & Speaker Gender Classification \cite{yang:2022} is a deep learning Jupyter Notebook project that classifies a person's gender based on an audio sample. \\ \hline
Income & 3 (1.02\%) & Assists in inclusive designs for bank services and home loan applications. & Inclusive Design Toolkit \cite{ontario:2022} provides documentation from the Government of Ontario, promoting diversity through services like income. \\ \hline
Sexual \newline \hfill Identity and \newline \hfill Orientation & 3 (1.02\%) & Creates an ontology for sexual orientation and gender.&Gender, Sex, and Sexual Orientation (GSSO) \cite{kronk:2022} is a web-based project for gender-related topics, such as sexual identity, orientation, and behaviour. \\ \hline
Social Media & 3 (1.02\%) & Profiles social media authors; Constructs social media followers. & A model, \texttt{gender\_prediction} \cite{gergely:2022}, within Jupyter Notebook that uses social media profiles to predict gender. \\ \hline
Analysis of \newline \hfill Gender Splits & 1 (0.34\%) & Analyses of gender distribution among TV and radio hosts and panel members. & The UK Panel Show Gender Breakdown \cite{lowe:2022} provides statistics of panel guests and hosts on UK TV shows. \\ \hline
%Sexual Identity & 1 (0.34\%) & Provides terms for gender and sexual identity; Supports paper on gender identity in social media.\\ \hline
\multicolumn{4}{c}{\cellcolor[gray]{0.8}\textbf{Institutional (n=32, 10.85\%)}} \\ \hline
Gender \newline \hfill Disparity & 14 (4.75\%) & Contains communities promoting equality in Tech; Analyses citations for gender balance; Promotes equality in legal representation.& PyLadies \cite{pyladies:2022} is a project owned by the PyLadies organisation that supports women in becoming active participants in the open-source community.\\ \hline
Inclusion & 9 (3.05\%) & Detects gender-inclusive language; Provides suggestions for mitigating exclusive language; Contains datasets on hate speech and gender-inclusive language. & Inclusive Design Toolkit \cite{ontario:2022} provides documentation and presentations from the Government of Ontario, Canada that promotes diversity through services, such as income and culture. \\ \hline
Gender \newline \hfill Diversity & 7 (2.37\%) & Promotes gender diversity in gaming and language; Contains a gender-balanced dataset for Artificial Intelligence. &  A project by GameHer \cite{gameher:2022}, a French association promoting gender diversity in the gaming community using tournaments and inclusive gaming environments.\\ \hline
Gender \newline \hfill Composition  & 2 (0.68\%) & Contains population information focusing on gender to promote positive change. & A project \cite{maize:2022} that supports a conference paper \cite{campenhout:2018} containing an algorithm that examines the effects of gender homophily in Uganda farmers when learning.\\ \hline
\multicolumn{4}{c}{\cellcolor[gray]{0.8}\textbf{Unrelated (n=25, 8.47\%)}} \\ \hline
Language & 19 (6.44\%) & Contains dictionaries, tools, and games on gendered languages and pronouns. & A French dictionary project \cite{belgacem:2022} in CSV format containing pronouns and verbs with gender.\\ \hline
Miscellaneous & 4 (1.35\%) & Provides ridership dataset for bike-sharing programs; Contains assignments, laser-cut prototyping, and hotel management system. & Citi Bike Analytics Jersey City 2020 \cite{amin:2022} examines data to better understand the growth rate of the service, which includes gender distribution of Citi Bike users.\\\hline
Gender Fields & 2 (0.68\%) & Provides gender data component for content management systems. & \texttt{craft3-fields} \cite{newism:2022} contains custom fields for Craft Content Management System that includes gender. \\ \hline
\multicolumn{4}{c}{\cellcolor[gray]{0.8}\textbf{Interactional (n=19, 6.44\%)}} \\ \hline
Gender bias & 18 (6.10\%) & Contains gender bias and debiasing tools; Lists gender bias papers; Analyses of gender bias using Google Translate. & A project, \texttt{gender-bias} \cite{marr:2022}, is a Python-based tool designed to identify and report gender-biased language in recommendation letters for medical professionals.\\ \hline
Gender Role & 1 (0.34\%) & Algorithm to support conference paper on gender homophily. & A project \cite{maize:2022} containing an algorithm that supports a conference paper \cite{campenhout:2018} that examines the effects of gender homophily for Uganda farmers when learning. \\ 
\end{tabular}
\end{table*}

The table contains 295 nodes to represent the coded projects. Some (n=98, 41.70\%) projects included more than one keyword, demonstrating their varying interests. Sixty (20.34\%) projects with more than one keyword were classified in multiple gender dimensions. For example, the project by Van Campenhout \cite{maize:2022} provides an algorithm used in a study \cite{campenhout:2018} that examines gender homophily \textemdash~the preference a person has for interacting with others of the same gender \cite{landiado:2016} \textemdash~during the training of farmers in Uganda. The study investigates gender composition and female and male trainers' roles in acquiring knowledge from farmers. The project contained the keywords \textit{``Gender Role''} and \textit{``Gender Composition''}, placing it in the two groups with the same names and classified within the gender social structure's \textit{Interactional} and \textit{Institutional} dimensions.  

The \textit{Individual} dimension contains the most (n=219, 74.24\%) projects. The majority (n=165, 55.59\%) are concentrated within the \textit{Detectors and Predictors} group. This group focuses on the detection and prediction of an individual's gender by using a person's name or by extracting information from audio and image files. For example, FaceLib \cite{ayoubi:2022} is a Python-based library that recognises an individual's facial expressions and gender through image files. Other projects in the \textit{Individual} dimension use an individual's attributes for \textit{Gender Classification} (n=20, 6.78\%), \textit{Income} (n=3, 1.02\%), and \textit{Sexual Identity and Orientation} (n=3, 1.02\%).

The \textit{Interactional} (n=19, 6.44\%) dimension, focusing on the cultural expectations placed on women and men, contains the least number of projects. For example, some projects concentrate on \textit{Gender Bias} (n=18, 6.10\%) that analyse gender-biased language. For instance, \texttt{gender-bias} \cite{marr:2022} is a project that contains a Python-based tool designed to identify and report gender-biased language in recommendation letters for the medical professions. Also, within the \textit{Gender Bias} group, a project \cite{zhao:2022} uses its \texttt{README} to list URL links to gender-related conferences and journal publications. The \textit{Gender Role} (n=1, 0.34\%) group contains one project \cite{maize:2022} that focuses on  society's expectations of how the genders behave, communicate, and conduct themselves. This previously mentioned project supports a study \cite{campenhout:2018} that examines how gender plays a role in the knowledge acquisition of Ugandan maize farmers educated by female and male trainers. 

The \textit{Institutional} (n=32, 10.85\%) dimension projects examine how culture influences gender-specific actions that emerge from regulations and organisational schemas. These include \textit{Gender Disparity} (n=14, 4.75\%) and \textit{Gender Diversity} (n=7, 2.37\%), which examine and promote gender equity and diversity in Tech. For example, the \textit{Gender Disparity} group contains a PyLadies \cite{pyladies:2022} project for the association's website maintained by their infrastructure team. PyLadies is an association that supports women in becoming active participants in the open-source community. The final category, \textit{Unrelated} (n=25, 8.47\%), contains 25 (8.47\%) projects that did not fit into the framework. Most (n=19, 6.44\%) of these projects focus on \textit{Language}, providing users with educational games and dictionaries. For example, one project \cite{belgacem:2022} contains a French dictionary that contains pronouns and verbs with their gender.

\begin{table*}[thp!]
\caption{Classification of Active Gender Bias GitHub Projects} \label{tbl:gender_bias_taxonomy_table}
\begin{tabular}{p{1.6cm}|p{2.8cm}|c|p{4.6cm}|p{5.4cm}}
\textbf{Types} & \textbf{Subtypes} & \textbf{Total (\%)} & \textbf{Definition}& \textbf{Example Project}\\ \hline
\multicolumn{5}{c}{\cellcolor[gray]{0.8}\textbf{Gender Bias Taxonomy \cite{doughman:2021}}} \\ \hline

\small{Occupational Bias} & \small{ $\bullet$ Gendered \newline\hfill Division of Labour \newline $\bullet$ Gendered Rules \& Duties} &  \small{13 (33.33\%)} & \small{Applies stereotypical gendered language to professions, such as referencing a role by a certain gender.} & \small{\texttt{gender-bias-BERT} \cite{bartl:2022} evaluates in English and German the gender bias in various professions.} \\ \hline

\small{Generic \newline \hfill Pronouns} & \small{$\bullet$ Generic He \newline$\bullet$ Generic She \newline$\bullet$ Gendered\newline \hfill Generic Man} & \small{8 (20.51\%)} & \small{Uses a generic pronoun with sex-indefinite antecedents.} &  \small{\texttt{gender-debias} \cite{saunders:2022} contains a debiasing dataset for ML training. Contains feminine and masculine German pronouns that change due to gender bias.}\\ \hline

\small{Semantics} & \small{$\bullet$ Metaphors \newline$\bullet$ Gendered \newline\hfill Attributes \newline$\bullet$ Old Sayings} & \small{8 (20.51\%)} & \small{Applies words to sentences in a demeaning way that alters their semantic meanings.} & \small{\texttt{gender-bias} \cite{marr:2022} is a Python-based tool designed to identify and report gender-biased language in recommendation letters for medical professions.} \\ \hline

\small{Sexism} & \small{$\bullet$ Hostile Sexism \newline$\bullet$ Benevolent Sexism}& \small{5 (12.82\%)} & \small{Applies words to create positive or negative expressions towards women and men depending on the context.} & \small{\texttt{Debiased-Chat} \cite{liu:2022} trains dialogue models to reduce gender bias, using the models to repose the pronouns that promote sexist expressions.} \\ \hline

\small{Exclusionary Bias} & \small{$\bullet$ Explicit Marking of Sex \newline$\bullet$ Gender-based \newline\hfill Neologisms \newline$\bullet$ Gendered Word \newline \hfill Ordering \& Division} & \small{2 (5.13\%)} & \small{Imposes gender on an unknown or neutral entity.} & \small{A project, \texttt{Double-Hard-Debias} \cite{wang:2022} supports the debiasing of word embeddings. Contains a dictionary that addresses exclusionary bias.} \\ \hline

\multicolumn{5}{c}{\cellcolor[gray]{0.8}\textbf{Emerging Codes}} \\ \hline

Imaging Bias && 2 (5.13\%) & Addresses biases that emerge during AI and ML image detection. & A project \cite{wang:2022A} that contains a framework that examines and mitigates various biases, including gender, within visual recognition tasks.\\ \hline

Discrimination && 1 (2.57\%) & Applies prejudicial treatment based on a person's gender or other background. & A project \cite{bhowmick:2022A} supporting a paper \cite{bhowmick:2022} analysing legal criminal cases in India to determine in-group bias within the judicial system. \\

\end{tabular}
\end{table*}

Our results show gender-related projects that promote change and raise awareness of the gender dynamics at the \textit{Institutional} and \textit{Interactional} dimensions. However, there is a limited (n=51, 17.29\%) number of projects focusing on these dimensions. A potential reason for these results might be the methods required to instigate change in these dimensions, which includes policies to promote diversity at educational institutions \cite{timmers:2010} and interventions to mitigate gender bias during the hiring process \cite{isaac:2019}. These methods to promote change and awareness might prominently use policies and documentation, whereas institutions and organisations might use other platforms and services outside of GitHub for their services. These methods might not require software applications and services to promote and disseminate change, which may explain the smaller representation of these projects on GitHub. However, more work is necessary to draw a conclusion.

\vspace{2mm}\hspace*{-4mm}\fbox{
\begin{minipage}{.97\linewidth}

\textbf{RQ2 Summary:}  The majority (n=219, 74.24\%) of the active gender-related projects focus on the individual, such as recognising and detecting a person's gender through input files. A small (n=19, 6.44\%) group of \textit{Interactional} projects focuses on the cultural expectations placed on women and men, which might require methods like policy to instigate and disseminate change. Organisations might not use GitHub to discuss gender policy, but more work is necessary to understand the limited focus on this dimension.

\end{minipage}}

\subsection{RQ3: What types of gender bias are actively explored within GitHub projects?} \label{subsection_rq_gender_bias}

The coding of all active gender-related projects identified 18 focusing on gender bias. Table \ref{tbl:gender_bias_taxonomy_table} shows these 18 projects classified in the gender bias taxonomy, organising the results by types followed by the emerging gender bias types identified during the coding process. The 18 projects produced 39 classification nodes, demonstrating that some projects focused on more than one gender bias type. For example, Table \ref{tbl:gender_bias_taxonomy_table} shows the \texttt{gender-bias} \cite{marr:2022} project \textemdash~a tool that examines letters of recommendation for the medical profession \textemdash~is classified under \textit{Semantics} and \textit{Occupational Bias}.

Table \ref{tbl:gender_bias_taxonomy_table} presents the coding for the gender bias taxonomy, including two emerging types, \textit{Imaging Bias} (n=2, 5.13\%) and \textit{Discrimination} (n=1, 2.57\%). These emerging types did not appear in the initial coding framework because the taxonomy focuses on gender bias in the English language. In contrast, \textit{Imaging Bias} focuses on biases during visual recognition tasks, and \textit{Discrimination} concentrates on a person's prejudicial treatment of a group based on their gender or other backgrounds, such as race. An exemplar project \cite{wang:2022A} for \textit{Imaging Bias} contains a framework that examines and mitigates various biases, including gender, within visual recognition tasks. For \textit{Discrimination}, the \texttt{gender-judicial-bias-india} project \cite{bhowmick:2022A} includes files necessary to replicate a study \cite{bhowmick:2022} evaluating the legal criminal cases in India for in-group bias. We observed the majority (n=11, 61.11\%) of the gender bias projects supporting research, such as \texttt{gender-bias-BERT} \cite{bartl:2022}, that supports a published thesis \cite{bartl:2020} that evaluates gender bias within different professional groups.

The results show the majority (n=13, 33.33\%) of the projects focused on \textit{Occupational Bias}, the stereotypical gendered language applied to societal roles. For example, \texttt{gender-bias-BERT} \cite{bartl:2022} is a project that contains a tool that evaluates the English and German languages for gender bias that exists in various professions, such as medical professions. The results show an equal number of projects (n=8, 20.51\%) within the \textit{Generic Pronouns} and \textit{Semantics} types. An exemplar \textit{Generic Pronouns} project is \texttt{gender-debias} \cite{saunders:2022}, which contains feminine and masculine German pronouns that change due to gender bias. An example of a \textit{Semantics} project is the previously mentioned \texttt{gender-bias} \cite{marr:2022} project that reports on gendered language within letters of recommendation for the medical profession.

The least applied gender bias type is \textit{Exclusionary Bias} (n=2, 5.13\%), a bias that assumes and assigns a gender to a neutral or unknown entity. For example, the \texttt{Double-Hard-Debias} \cite{wang:2022} project focuses on debiasing \textit{word embeddings}, a natural language processing (NLP) concept that analyses text through word representation.

The \textit{Imaging Bias} (n=2, 5.13\%) type contains projects considering the stereotypes and inequities that emerge from visualisation tools. However, we observe from our findings a disparity between those projects (n=2) mitigating imaging bias and (n=164) projects focusing on visual recognition through gender (See Section \ref{subsection_rq4}). Perhaps some 164 projects are considering imaging bias, but additional analysis is necessary to draw a conclusion.

\vspace{2mm}\hspace*{-4mm}\fbox{
\begin{minipage}{.97\linewidth}

\textbf{RQ3 Summary:} We were able to code most (n=15, 83.33\%) of the gender bias projects within the taxonomy. Three projects addressed gender bias behaviours within visualisation tools and discrimination. \textit{Occupational Bias} is the most common type, while \textit{Exclusionary Bias} is the least applied. Though the two emerging types contained a small (n=3, 7.70\%) number of gender bias projects, it demonstrates the open-source community is considering aspects of gender bias within social behaviours and mitigating bias within the software. 

\end{minipage}}

\section{Implications} \label{section_implications}

\textbf{To Educators:}
We encourage educators to teach the ethical implications of gender detection because our findings show a strong interest in automated methods for identifying an individual's gender. \textit{Nature Portfolio} surveyed artificial intelligence (AI) experts, collecting their opinions on facial-recognition research \cite{noorden:2020}. The results showed some experts had ethical concerns in the AI field. Teaching students the implications of unethical projects might help address these concerns. Educators can also help raise students' awareness of interactional topics, such as gender bias, to potentially get them involved in open-source projects that address these topics. This might motivate students to get involved in ethical practices, such as research labs \cite{kumar:2018} that use machine and deep learning technologies to identify hate speech. Educators can also minimise students' exposure to gender-biased software. For example, a checklist exists for K-12 programs \cite{bhargava:2002} for educators to help identify gender biases in software. The checklist can assist educators in making informed decisions on adopting more inclusive software.

%We observed from our findings  that some of the gender-related projects supporting published research. 

\textbf{To Open-Source Community:} The open-source community can increase awareness of the gender-related work being conducted in their community. One way to raise awareness is through the GitHub collections, a curated list of projects posted on GitHub that highlight the work being done in communities and industries on various topics and interests.\footnote{https://github.com/collections} Showcasing these projects as a collection could encourage more members of the open-source community to get involved in these gender-related projects.

\textbf{To Researchers:} The research community may be addressing gender-related topics outside of the GitHub platform in the \textit{Interactional} dimension. We encourage researchers to make available their work on promoting diversity, equity, and inclusion on open-source platforms. The presence of this work on open-source platforms might encourage the community to create and adapt projects to address gender-related topics, especially areas with a narrow focus, like \textit{Imaging Bias}.

\section{Threats to Validity} \label{section_threats_to_validity}

There are limitations to this research. The first relates to content validity since our study's reliability depends on the projects identified by the pipeline. We cannot say that the pipeline identified all gender-related projects; however, we used the GitHub REST API to validate the collected data to strengthen the results. Another threat is data consistency. To reduce the margin of variability in the qualitative results, we applied the approaches Silverman \cite{silverman:2009} suggested to improve reliability. The approaches include constant data comparison, in which the authors discuss the coding criteria, process, and results until agreeing.

\section{Conclusion and Future Work} \label{section_conclusion}

% A section of the pipeline reporting on Google Scholar publications could not report findings due to issues parsing data.

%active gender-related GitHub projects to better understand the open-source community's efforts and interest in gender-related topics through the development of open-source projects. A

The motivation for this paper is to understand the open-source community's interest in gender-related topics by evaluating their active projects on GitHub. We used a pipeline to identify the active gender-related GitHub projects, classifying them twofold, first identifying the active projects within a gender social structure and then classifying projects within a gender bias taxonomy. The results show most gender-related projects focus on the individual, performing gender detection and recognition. We observed a limited focus on interactional topics, such as projects addressing gender bias. Within gender bias, these projects primarily focus on occupational bias, with a small representation focusing on discrimination.

%For all the gender-related projects, the results show the composition of project owners is similar to other active GitHub projects. However, the interest from the open-source community is lower for these
%For textual analysis, pre-processing the data was required for accurate reporting. Having an existing pipeline can help researchers mine projects faster and reliably, but future work can strengthen the pipeline by including selection criteria to identify active projects, integrate recommendations from the mining software repositories (MSR) community to remove biased results. 

Future research can investigate the dearth of interactional projects to understand better the limited focus on detecting gendered behaviours. As Section \ref{section_implications} mentions, the education, open-source, and research communities can raise awareness of projects focusing on interactional topics, helping promote more project development in this area. There is a possibility GitHub projects focus on interactional topics, but these projects might not specify their intentions through keywords and descriptions. Future research can perform additional queries to identify other gender-related projects and can analyse more projects in-depth to identify their purpose better.

%and potentially focus on other interactional aspects, such as interactions, communications, and cooperation between the genders. 

%providing the number of contributors to these projects, enabling them to view the interests the open-source software community has in this area.

\bibliographystyle{IEEEtran}
\balance
\bibliography{sample-base}

\end{document}